\documentclass[preprintnumbers,amsmath,amssymb,superscriptaddress]{revtex4-2}

\usepackage{mathrsfs}
\usepackage{amssymb}
\usepackage{graphicx}
\usepackage{dcolumn}
\usepackage{bm}
\usepackage{textcomp}
\usepackage{amsmath}
\usepackage[titletoc]{appendix}
\usepackage{accents}
\usepackage{ntheorem}
\usepackage{color}

\newcommand\ket[1]{\ensuremath{|#1\rangle}}
\newcommand\bra[1]{\ensuremath{\langle#1|}}
\newcommand\iprod[2]{\ensuremath{\langle#1|#2\rangle}}
\newcommand\oprod[2]{\ensuremath{|#1\rangle\langle#2|}}
\newcommand\mean[1]{\ensuremath{\langle #1\rangle}}
\newcommand\tr{\mathop{\rm tr}\nolimits}
\newcounter{RomanNumber}

\makeatletter
\def\widebar{\accentset{{\cc@style\underline{\mskip10mu}}}}
\def\Widebar{\accentset{{\cc@style\underline{\mskip8mu}}}}
\makeatother

\begin{document}

\title{Side-channel-secure quantum key distribution}

\author{Cong Jiang}
\affiliation{Jinan Institute of Quantum Technology, Jinan, Shandong 250101, China}
\affiliation{State Key Laboratory of Low Dimensional Quantum Physics, Department of Physics, Tsinghua University, Beijing 100084, China}

\author{Xiao-Long Hu}
\affiliation{School of Physics, State Key Laboratory of Optoelectronic Materials and Technologies, Sun Yat-sen University, Guangzhou 510275, China}

\author{Zong-Wen Yu}
\affiliation{Data Communication Science and Technology Research Institute, Beijing 100191, China}

\author{Xiang-Bin Wang}\email{Corresponding author: xbwang@mail.tsinghua.edu.cn}
\affiliation{Jinan Institute of Quantum Technology, Jinan, Shandong 250101, China}
\affiliation{State Key Laboratory of Low Dimensional Quantum Physics, Department of Physics, Tsinghua University, Beijing 100084, China}
\affiliation{Frontier Science Center for Quantum Information, Beijing, China}
\affiliation{Shanghai Branch, CAS Center for Excellence and Synergetic Innovation Center in Quantum Information and Quantum Physics, University of Science and Technology of China, Shanghai 201315, China}
\affiliation{ Shenzhen Institute for Quantum Science and Engineering, and Physics Department, Southern University of Science and Technology, Shenzhen 518055, China}

\begin{abstract}
We present a result of side-channel-secure (SCS) quantum key distribution (QKD) under fully realistic conditions. Our result is not only measurement-device independent but also effective with imperfect (and unstable) source devices including imperfect vacuum and imperfect coherent-state source. 
Applying the virtual mapping idea, we present a general security proof under whatever out-side-lab attack, including whatever side-channel coherent attack. As a by- product, we also present  an improved method for SCS protocols which can raise the key rate by 1-2 orders of magnitude. Using these results, we obtain a non-asymptotic key rate which is instantly useful with full realistic conditions.  
\end{abstract}


\maketitle
{\em Introduction.} Since the first quantum key distribution (QKD) protocol---BB84 was proposed in 1984~\cite{bennett1984quantum,gisin2002quantum,xu2020secure,pirandola2020advances,scarani2009security,
lo2012measurement,braunstein2012side,wang2019practical}, the practical security\cite{hwang2003quantum,wang2005beating,lo2005decoy,lo2012measurement,braunstein2012side}, the key rate and secure distance\cite{lu2018overcoming,wang2018twin} have been greatly improved so far. Especially, the measurement-device-independent(MDI)-QKD protocol~\cite{lo2012measurement,braunstein2012side} and the twin-field QKD~\cite{lu2018overcoming,wang2018twin} are immune to all attacks to the detectors. Yet, the source side channels are threatening the security of practical QKD systems. Recently, the side-channel-secure (SCS) QKD protocol was proposed~\cite{wang2019practical} for a security against all out-side-lab attacks to the emitted states side channels while keeping the MDI security for detectors, if there are perfect vacuum states for use and the intensity of coherent states is upper bounded. More recently, a 50 km fiber SCS QKD experiment was done~\cite{zhang2022experimental}, showing the potential of SCS-QKD protocol for practical applications in the future. 

However, given that the most crucial problems here, the imperfect vacuum and security under coherent attack with finite number of pulses still open, there is still a difficult gap between the potential and real security against side channels. Although there are useful discussions on the issue of imperfect vacuum ~\cite{zhang2022experimental,jiang2023side}, however, the proposed solution assumed either stable source states which are not accessible in practice or using different states at different time windows which can not meet the request of permutation invariant~\cite{christandl2009postselection} and hence the most important issue,  the practical security under coherent attack with finite data size turned out to be unsolvable. 

Here in this work, by the delicate method of constructing a map capable of mapping the perfect protocol to the real protocol, we thoroughly solve all open problems for SCS protocol under realistic conditions and with our result here the side-channel-secure QKD comes true immediately with existing technologies. It only needs the practically accessible imperfect coherent states and imperfect vacuum, with crude bound values of the light intensity, which are easy manipulate in practice. Notably, it does not need any side channel information of those imperfect states.

In what follows, we shall first present the general security proof for real SCS protocols. After that we present key rate calculation for real SCS protocols with imperfect source devices and finite numer of pulses. As a by-product, we also present improved SCS protocols which can raise the key rate by one magnitude order.
    
{\em General security proof of real SCS protocols by mapping.}

The main idea here is to mathematically find a map $\mathcal M$ which can map the perfect states of a virtual perfect protocol into the imperfect states which are actually used in a real protocol. If such a map in principle exists, the real protocol using imperfect states must be secure if the virtual protocol using perfect states is secure. Otherwise, Eve can attack the perfect protocol effectively: she can first use map $\mathcal M$ to change the perfect states into imperfect states used in the real protocol and then apply the attacking method which can attack the real protocol effectively. For clarity, let's first recall Theorem 1 of Ref.~\cite{wang2019practical} presented below as a lemma:

Lemma. If there is in principle a map $\mathcal{M}$ which can map the (virtual) perfect protocol $\mathcal P$ into the real protocol $\mathcal R$, then the real protocol $\mathcal R$ must be secure if the perfect protocol $\mathcal P$ is secure and we can regard the real protocol as perfect protocol in key rate calculation.

Remark: If Eve can attack the real protocol by $\mathcal G$, she can also attack the virtual protocol equivalently by $\mathcal {G\cdot M}$. Alternatively, we can regard the real protocol as the virtual perfect protocol with a specific channel whose first action is to map those perfect states into  imperfect real states. If a perfect protocol is proven to be secure, it must be also secure under this specific channel.

In the existing SCS protocols~\cite{wang2019practical,jiang2023side} ,there are two sources for each of Alice and Bob. At Alice’s (Bob’s) side, she (he) has a coherent-state source $x_A$ ($x_B$) and a vacuum source $o_A$ ($o_B$).

1. Virtual perfect SCS protocol $\mathcal P$ and real SCS protocol $\mathcal R$\\
A. Virtual perfect SCS protocol $\mathcal P$:\\
In a virtual perfect SCS protocol $\mathcal P$, all source devices are perfect. At every time window, the coherent-state source $x_A$ ($x_B$)  emits a perfect coherent state $|\mu_A\rangle$ ( $|\mu_B\rangle$) as it is supposed to, while the vacuum sources will emit a perfect vacuum. These perfect coherent states and perfect vacuums in the perfect protocol $\mathcal P$ have no side channel imperfection and below we write them in the photon-number space only:
\begin{equation}\label{virtual_statep1}
\begin{split}
&\ket{\mu_A} = e^{-\mu_A/2}  \ket{0} +\sqrt{1-e^{-\mu_A}}\ket{q_A},\\
&\ket{\mu_B}=e^{-\mu_B/2}  \ket{0} + \sqrt{1-e^{-\mu_B}}\ket{q_B}.\\
\end{split}
\end{equation}
where $\sqrt{1-e^{-\mu_A}}\ket {q_A}=\sum_{k=1}^\infty \frac{\mu_A^{k/2}}{\sqrt k!}\ket{k}$ and $\sqrt{1-e^{-\mu_B}} \ket {q_B}=\sum_{k=1}^\infty \frac{\mu_B^{k/2}}{\sqrt k!}\ket{k}$. If Alice (Bob) chooses the vacuum source $o_A$ ($o_B$), she (he) prepares the perfect vacuum state $\ket{0}$.
\\
B. Real SCS protocol $\mathcal R$:\\
 In a real SCS protocol $\mathcal R$, the source devices are imperfect. The source states have imperfections in the whole space including both the imperfection in side channel space such as frequency spectrum, emission time and so on, and imperfection in photon number space. Also, the sources are in general unstable.
Most generally, at time window
$i$, the whole-space states from the imperfect coherent-state sources $x_A$ and $x_B$ can be written in the following form:
\begin{equation}\label{real_statep2}
\begin{split}
&\ket{\psi _{Ai}}=\sqrt {a_{0i}}  \ket{0}+\sqrt{1- a_{0i} } \ket{\tilde \psi_{Ai}},\\
&\ket{\psi _{Bi}}=\sqrt {b_{0i}} \ket{0}+\sqrt {1- b_{0i}} \ket{\tilde \psi_{Bi}}.
\end{split}
\end{equation}
Here $\ket{\tilde \psi_{Ai}}$ ($\ket{\tilde \psi_{Bi}}$) is a whole-space state containing at least 1 photons. That is to say, the each of whole-space states are written in the linear superposition of vacuum and non-vacuum. Obviously, we have 
\begin{equation}
\iprod{0}{\tilde \psi_{Ai}}=\iprod{0}{\tilde \psi_{Bi}}=0.
\end{equation}

It is crucially important that, in the real protocol, the vacuum source is also imperfect. Say, at any  time window $i$ of the real protocol, if Alice (Bob) chooses the source $o_A$ ($o_B$), she (he) actually prepares the state:
\begin{equation}\label{real_state1}
\begin{split}
&\ket{v_{Ai}}=\sqrt{a_{v0i}}\ket{0}+\sqrt{1-a_{v0i}}\ket{\tilde{ \phi}_{Ai}},\\
(&\ket{v_{B{i}}}=\sqrt{b_{v0i}}\ket{0}+\sqrt{1-b_{v0i}}\ket{\tilde{\phi}_{Bi}}).
\end{split}
\end{equation}
Here $\ket{0}$ is the vacuum state and state $\ket{\tilde {\phi}_{Ai}}$ ($\ket{\tilde {\phi}_{Bi}}$) is a whole-space states containing at least 1 photons. Obviously, we have 
\begin{equation}
\iprod{0}{\tilde \phi_{Ai}}=\iprod{0}{\tilde \phi_{Bi}}=0.
\end{equation}

In the formulas for imperfect states above, we have set all coefficients there to be positive real numbers rather than complex numbers because the non-vacuum states $\ket{\tilde \psi_{Ai}}$ , $ \ket{\tilde \psi_{Bi}}$ in the linear superposition can contain whatever complex factors themselves. 

In the imperfect states above, {\em they} (Alice and Bob) do not have detailed information of the whole-space state, but they may know the bounds of the vacuum amplitude values  such as  $\sqrt{a_{v0i}}$ in the superposition.

2. Mapping from virtual perfect protocol $\mathcal P$ to real protocol $\mathcal R$: attenuation and unitary transformation.

We now present our central result of this work: the map $\mathcal{M}$ that maps perfect states of protocol $\mathcal P$ into imperfect states of real protocol $\mathcal R$ do exist. The map $\mathcal M$ consists of two steps: attenuation and unitary transformation. 

In general, there isn't a unitary transformation that directly maps the perfect  states of Eq. (\ref{virtual_statep1}) in virtual perfect protocol to the imperfect states of Eqs. (\ref{real_statep2},\ref{real_state1}) in real protocol: the fidelity of two states in the virtual perfect protocol is a constant while the fidelity of the two states in the real protocol can be different at different time windows. The constant fidelity value in the virtual perfect protocol cannot be equal to the time-dependent fidelity values in the real protocol. However, we can take a two-step map beyond unitary transformation to obtain the real states from perfect states. The map is in general different at different time windows.

For simplicity,  below we shall only consider the source states at Alice’s side as the states at Bob’s side can obviously be treated in a similar way.

A. Attenuation $\hat D_A (\xi)$ to the perfect states.
\\
We first consider the attenuation operation to the source states of virtual protocol $\mathcal P$. 
Denoting the attenuation of the time window $i$ by $\hat D_A (\xi_{Ai}) $ where $0\le \xi_{Ai}\le 1$ is the transmittance of the attenuator. Explicitly, we have 
\begin{equation}\label{attad}\begin{split}
&\ket{t_{0i}}= \hat D_A (\xi_{Ai}) \ket{0}=\ket{0},\ket{t_{ci}}=\hat D_A (\xi_{Ai}) \ket{\mu_A} \\
&= \ket{r_{Ai} =\xi_{Ai}\cdot \mu_A}
=\sum_{k=0} \frac{e^{-r_{Ai}/2}(r_{Ai})^{k/2}}{\sqrt{k!}}\ket{k}.
\end{split}
\end{equation}  
Surely, we can set whatever value of $0\le\xi_{Ai}\le 1$ in this process, and it can be different at different time windows. We have the following fact:\\
Fact 1:  Mathematically, there exists attenuation operation to map the coherent state $\ket{\mu_A}$ to whatever coherent state $\ket{r_{Ai}}$ provided that $\mu_A \ge r_{Ai} \ge 0$.

B.Unitary transformation $\mathcal U_{Ai}$

After attenuation, the source states of perfect protocol are changed into
coherent state $\ket{r_{Ai}}$ and vacuum $\ket{0}$. They have the fidelity
\begin{equation}
F_{Ai} =\left| \iprod{0}{r_{Ai}}\right|^2
= e^{-r_{Ai}}.
\end{equation}
Obviously, for any $i$, we have the following range $S_P$ for fidelity  $F_{Ai}\in S_P$:
\begin{equation}\label{attp}
    S_P= [e^{-\mu_A}, 1].
\end{equation}

On the other hand, the fidelity of two states in the real protocol given by Eqs.(\ref{real_statep2},\ref{real_state1})  are
\begin{equation}
\begin{split}
    &f_{Ai} = |\iprod{v_{Ai}}{\psi_{Ai}}|^2 \\
    &= \left|\sqrt{a_{0i}a_{v0i}}+\sqrt{(1-a_{0i})(1-a_{v0i})}\iprod{\tilde\phi_{Ai}}{\tilde\psi_{Ai}}\right|^2.
    \end{split}
\end{equation}
Obviously, we have
\begin{equation}
\begin{split}
&    \left|\sqrt{a_{0i}a_{v0i}}-\sqrt{(1-a_{0i})(1-a_{v0i})}\right|^2 \le \\
 &   f_{Ai} \le  \left|\sqrt{a_{0i}a_{v0i}}+\sqrt{(1-a_{0i})(1-a_{v0i})}\right|^2.
    \end{split}
\end{equation}
Given the bounds of
\begin{equation}\label{boundA}
    a_{0i} \ge  {a_0}\ge 0.5;\;
    a_{v0i} \ge  {a_{v0}}\ge 0.5.
\end{equation}
for any $i$.
we obtain the following range $S_R$ for fidelity  $f_{Ai}\in S_R$:
\begin{equation}\label{attr}
    S_R = \left[\left|\sqrt{ a_0 \cdot a_{v0}}-\sqrt{(1-{a_{0}})(1-{a_{v0}})}\right|^2, 1\right]
\end{equation}
If the range $S_P$ in Eq.(\ref{attp}) covers the whole range $S_R$ in Eq.(\ref{attr}), then there must exist an attenuation process by which  the fidelity of two states  $\ket{t_{0i}}$ and $\ket{t_{ci}}$ in Eq.(\ref{attad}) exactly equals to the fidelity of the two real states at time window $i$.
To make sure that the range $S_P$ in Eq.(\ref{attp}) covers the whole range $S_R$, we only need
\begin{equation}\label{condition1}
 e^{-\mu_A} \le    \left|\sqrt{ a_{0} \cdot a_{v0}}-\sqrt{(1-{a_{0}})(1-{a_{v0}})}\right|^2 
\end{equation}
Therefore, we conclude that if  Eq.(\ref{condition1}) is satisfied, there must be an appropriate transmittance value $\xi_{Ai}$ in the attenuation $\hat D_A$ after which the states of the perfect protocol 
$\{\oprod{0}{0}, \oprod{\mu_{A}}{\mu_{A}}\}$ can be mapped to the imperfect whole-space states $\{\oprod{v_{ai}}{v_{ai}}, \oprod{\psi_{Ai}}{\psi_{Ai}}\}$ in the real protocol by a unitary transformation. 

Obviously, we can take the similar conclusion to states at Bob's side, with the condition
\begin{equation}\label{condition2}
  e^{-\mu_B} \le    \left|\sqrt{ b_{0} \cdot b_{v0}}-\sqrt{(1-{b_{0}})(1-{b_{v0}})}\right|^2   
\end{equation}
where for any $i$
\begin{equation}\label{boundB}
    b_{0i} \ge  {b_0}\ge 0.5;\;    b_{v0i} \ge  {b_{v0}}\ge 0.5.
\end{equation}

Therefore we have the following theorem:
\\Theorem 1. Given conditions of Eqs.(\ref{condition1},\ref{condition2}), there exists a map  which for sure to map the virtual SCS protocol $\mathcal P$ into the real SCS protocol $\mathcal R$.

This theorem has actually made a complete security proof for the real SCS protocol $R$, and we can easily carry out the real SCS protocol in the following way: 

{\bf Major conclusion}: Given any real SCS protocol $\mathcal R$, using the bound values of $a_0,a_{v0},b_0,b_{v0}$ as requested by Eqs.~(\ref{boundA},\ref{boundB}), we can calculate the corresponding values of $\mu_A,\mu_B$ by constraints Eqs.~(\ref{condition1},\ref{condition2}). As we have already show the security of perfect SCS protocol $\mathcal P$ in prior art literatures\cite{wang2019practical,jiang2023side}, we can simply obtain a secure key rate for the real SCS protocol just by regarding it as a perfect protocol using perfect coherent states $\{\ket{\mu_A}, \ket{\mu_B}\}$ and perfect vacuum. Since all states in the perfect protocol are perfect and stable, the secure key with finite data size can be obtained for the real protocol because the protocol is permutation invariant. The result is secure under whatever coherent side-channel attacks outside Alice's and Bob's labs. 

Remark: Surely, our method above is also robust to state imperfections in photon number space. We do not request any specific photon number distribution such as Poisson distribution for the real sources. We only need lower bounds of amplitude of vacuum term in Eqs.(\ref{real_statep2},\ref{real_state1}).

\begin{figure}
\centering
\includegraphics[width=8cm]{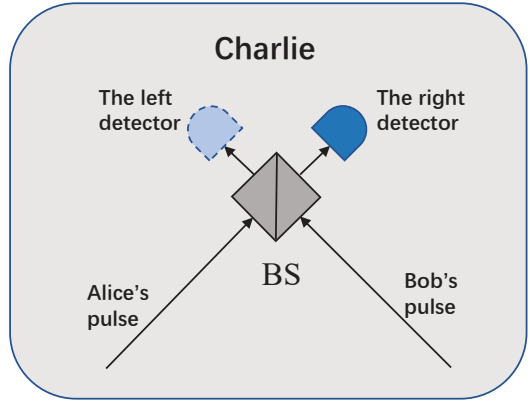}
\caption{The model of Charlie's measurement equipment. BS: The 50:50 beam splitter.}\label{equip}
\end{figure}

{\em Side-channel-secure key under fully realistic conditions}\\
Let's first recall SCS protocol below: In the $i$-th time window, Alice (Bob) randomly prepare a pulse from sources $o_A$ or $x_A$ ($o_B$ or $x_B$) with probabilities $p_0$ and $p_x=1-p_0$, and she (he) puts down a bit value 1 (0) if she (he) decides to use source $x_A$ ($x_B$), and a bit value 0 (1) if she (he) decides to use source $o_A$ ($o_B$). Then \emph{they} (Alice and Bob) send the prepared pulses to Charlie shown in Fig.~\ref{equip}. {\em They}  repeat the above process for $N$ times. Charlie announces those effective time windows to Alice and Bob. {\em They} process data from effective time windows, estimating bit-flip error and phase-flip error, and calculate the final key. 

For clarify, we have the following definitions:
\\ \textbf{The $\mathcal{O}$ window:}  a time window when Alice chooses the source $o_A$ and Bob chooses the source $o_B$.
\\ \textbf{The $\mathcal{B}$ window:} a time window when Alice chooses the source $x_A$ and Bob chooses the source $x_B$.
\\ \textbf{The $\mathcal{Z}$ window:} a time window when Alice choose the source $o_A$ and Bob chooses the source $x_B$, or Alice choose the source $x_A$ and Bob chooses the source $o_B$. The bits of $\mathcal{Z}$ windows are untagged bits.
\\ ${n}_{\zeta}\mbox{ }(\zeta=\mathcal{O},\mathcal{B},\mathcal{Z})$: the the number of effective $\zeta$ windows.

Our improved method for Charlie's action. Here Charlie is supposed to take phase compensation at all kinds of time windows and it will cause significant consequence for $\mathcal{B}$ windows. In SCS protocol, most of the bit-flips come from $\mathcal{B}$ windows, where Alice and Bob have chosen to send out a state from sources $x_A$ and $x_B$ respectively. In the improved method presented here, Charlie is supposed to take phase compensation so that in a $\mathcal{B}$ windows, after passing through the BS, (most of) the light energy is supposed to be at the left side of the BS. If the right side detector clicks and the left side detector does not click, Charlie announces that it is an effective window to Alice and Bob. In this way, heralding of any $\mathcal{B}$ windows is very unlikely. In a $\mathcal{Z}$ window or $\mathcal{O}$ window, Charlie’s phase compensation does not change the clicking probability of each detector. This greatly reduces the QBER and hence this method can improve the sending probability. As shown in the method part, there are variants of the improved method here.

{\em They} (Alice and Bob) use data from effective windows for the raw bits. In the data post-processing, {\em they} first take error correction with the input value of bit-flip rate obtained by error test with their raw bits.

For the collective attack, we can calculate the upper bound of the phase-flip error rate $\bar{e}^{ph}$ according the values of $n_{\zeta}$, and the details are shown in the method part. Then, Alice and Bob can calculate the secure final key rate under collective attack by~\cite{scarani2008security,sheridan2010finite}
\begin{equation}\label{keyrate}
\begin{split}
&R_{col}=\frac{1}{N}\left\{n_{\mathcal{Z}}[1-H(\bar{e}^{ph})]-leak_{EC}-\log_2\frac{2}{\varepsilon_{cor}}\right .\\
&\left .-2\log_2\frac{1}{\varepsilon_{PA}}-(d+3)\sqrt{n_{\mathcal{Z}}\log_2\frac{2}{\bar{\varepsilon}}} \right\}.
\end{split}
\end{equation}
Here, $leak_{EC}$ is the amount of information leakage during the error correction process and generally $leak_{EC}=fM_sH(E_{\mathcal{Z}})$ where $f$ is the error correction inefficiency, and $M_s=n_{\mathcal{O}}+n_{\mathcal{B}}+n_{\mathcal{Z}}$ is the number of raw keys, and $E_{\mathcal{Z}}$ is the bit-flip error rate of the raw key strings; $H(x)=-x\log_2x-(1-x)\log_2(1-x)$ is the Shannon entropy; $\varepsilon_{cor}$ is the failure probability of the error correction; $\varepsilon_{PA}$ is the failure probability of the privacy amplification; $\bar{\varepsilon}$ is the coefficient of measuring the accuracy of estimating the smooth min-entropy; $d$ is the dimension of the local states shared by Alice and Bob, and $d=8$ in SCS protocol. 

While Alice and Bob perform the privacy amplification process according to Eq.~\eqref{keyrate}, the protocol is $\varepsilon_{col}$-secure under collective attack \cite{sheridan2010finite}, and
\begin{equation}
\varepsilon_{col}=\varepsilon_{cor}+\bar{\varepsilon}+\varepsilon_{PA}+n_{PE}\epsilon,
\end{equation}
where $\epsilon$ is the probability that the real value of a parameter lies outside of the chosen fluctuation range, and $n_{PE}$ is the number of how many parameters are needed to estimate. In this protocol, $n_{PE}=3$.

Applying the post-selection technique~\cite{christandl2009postselection}, by shorten $2(d^2-1)\log_2(N+1)$ bits of the final key distilled under collective attack, we can get secure final keys against the coherent attack. Finally, we can get the key rate formula under coherent attack by 
\begin{equation}\label{rcoh1}
R_{coh}=R_{col}-\frac{2(d^2-1)\log_2 (N+1)}{N},
\end{equation} 
and the corresponding security coefficient under coherent attack is~\cite{christandl2009postselection} 
\begin{equation}\label{coh}
\varepsilon_{coh}=\varepsilon_{col}(N+1)^{d^2-1}.
\end{equation}

\emph{Numerical simulation}

We use the linear model to simulate the observed values. The experiment parameters are listed in Table.~\ref{exproperty}.

\begin{table*}[htbp]
\centering
\begin{tabular}{cccccc}
\hline
$p_d$& $e_d$ &$\eta_d$ & $f$ & $\alpha_f $ & \textcolor{red}{$\varepsilon_{coh}$} \\
\hline
$1.0\times10^{-9}$& {$4\%$}  & {$30.0\%$} & $1.1$ & $0.2$ & $10^{-10}$\\ 
\hline
\end{tabular}
\caption{List of experimental parameters used in numerical simulations. Here $p_d$ is the dark counting rate per pulse of Charlie's detectors; $e_d$ is the misalignment-error probability; $\eta_d$ is the detection efficiency of Charlie's detectors; $f$ is the error correction inefficiency; $\alpha_f$ is the fiber loss coefficient ($dB/km$).}\label{exproperty}
\end{table*}

\begin{figure}
\centering
\includegraphics[width=9cm]{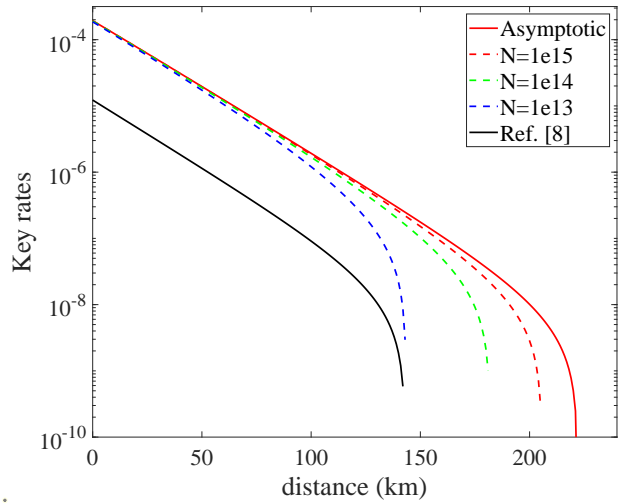}
\caption{The key rate of the improved method with different block size. We set \textcolor{red}{$a_{v0}=b_{v0}=1-10^{-8}$ for sources $o_A,o_B$} while the intensities of sources $x_A,x_B$ fluctuates in $\pm 10\%$.}\label{finite}
\end{figure}

Figure~\ref{finite} is the key rate of the improved method with different block size, where we set \textcolor{red}{$a_{v0}=b_{v0}=1-10^{-8}$ for sources $o_A,o_B$} while the intensities of sources $x_A,x_B$ fluctuates in $\pm 10\%$. The experiment parameters are listed in Table.~\ref{exproperty}. For the asymptotic case, comparing with the results of Ref.~\cite{wang2019practical}, the key rate can be improved by 1-2 orders of magnitude through our improved method. For the non-asymptotic case, in the short distances (less than 100 km), the key rates are less affected by the finite-key size effect and block sizes, but the key rate drops sharply as the block size decreases in the long distance. But still, the key rate of the improved method in the non-asymptotic case is much higher than that of the original protocol in the asymptotic case.

\emph{Conclusion.} By proving the existence of the mapping from the ideal protocol to the real protocol, we present non-asymptotic result of SCS protocol with our improved method. Our security proof is applicable to all source state imperfections in the whole space, including imperfections in both side channel space and photon number space provided that the lower bounds of vacuum amplitude are known. Our result here is secure under whatever out-side-lab side-channel coherent attack. The numerical results show that the key rate can be improved by 1-2 orders of magnitude through our improved method, and increase the farthest distance to nearly 200 km. With this work, the side-channel-secure QKD with fully realistic conditions comes true immediately.

\textbf{Funding.} This work was supported by  National Natural Science Foundation of China Grant Nos. 12174215, 11974204, 12104184, and 12147107; Shandong Provincial Natural Science Foundation Grant No. ZR2021LLZ007; Key R$\&$D Plan of Shandong Province Grant No. 2021ZDPT01; Open Research Fund Program of the State Key Laboratory of Low-Dimensional Quantum Physics Grant No. KF202110; Ministration of Science and Technology of China through The National Key Research and Development Program of China Grant No. 2020YFA0309701; Leading Talents of Quancheng Industry.

\bibliography{refs-jiang.bib}

\setcounter{section}{0}
\setcounter{equation}{0}
\section*{Method}
\renewcommand\theequation{S\arabic{equation}}
\subsection{Calculation of phase error rate of SCS protocol in the non-asymptotic case}
Most of the technique used here is the same with those of Refs.~\cite{wang2019practical,hu2022universal,jiang2023side}, for the completeness of this work, we show the security proof as follows. 

As shown in the main text, the real protocol can be regarded as a virtual perfect protocol which can be mapped into the real protocol. Explicitly, as shown in Eqs. (13,14) in the main text, given the lower bounds of $a_{v0i} \ge a_{v0}, a_{0i} \ge a_0, b_{v0i} \ge b_{v0}, b_{0i} \ge b_0$ for all ${i}$ in the real protocol, we shall regard it as a perfect protocol with perfect vacuum states and perfect coherent states of $|\mu_A\rangle(|\mu_B\rangle)$ at Alice's (Bob's) side with exact values 
\begin{align}
 &e^{-\mu_A} =    \left|\sqrt{ a_{0} \cdot a_{v0}}
    -\sqrt{(1-{a_{0}})(1-{a_{v0}})}\right|^2 ,\\
  &e^{-\mu_B} =    \left|\sqrt{ b_{0} \cdot b_{v0}}
    -\sqrt{(1-{b_{0}})(1-{b_{v0}})}\right|^2.   
\end{align}
for all time windows. In what follows we calculate the phase error and key rate of such a perfect protocol.

1. State preparation. In each time window, Alice and Bob prepare the following joint state:
\begin{equation}
\ket{\phi}=p_0\ket{00}_S \otimes\ket{2}_{I}\otimes \ket{\mathcal{O}}_J+p_x\ket{\mu_A\mu_B}_S\otimes\ket{2}_{I} \otimes\ket{\mathcal{B}}_J+\sqrt{2p_0p_x}\ket{\Psi}\otimes \ket{\mathcal{Z}}_J,
\end{equation} 
where
\begin{equation} \label{pshi}
\ket{\Psi}=\frac{1}{\sqrt{2}}(\ket{0\mu_B}_S\otimes \ket{01}_{{I}}+\ket{\mu_A0}_S\otimes \ket{10}_{{I}}),
\end{equation}
and $\iprod{01}{2}_I=\iprod{10}{2}_I=0$. Subsystems $J$ and $I$ are stored in Alice's and Bob's labs, and the subsystem $S$ is sent to Charlie. In the absence of misunderstanding, we will omit subtitles $S,{I},J$.

2. Detection. Alice and Bob send the subsystem $S$ to Charlie and repeat the above precess for $N$ times. Charlie performs interaction with the states $\ket{\phi}^{\otimes N}$ and stores all the measurement results (whether it is an effective window or not; for simplicity, we also call it clicking or not) in his instrument $\mathcal{L}$, which can be observed by Alice and Bob.

Remark: since the SCS QKD protocol is MDI, we can always assume the measurement station is controlled by Eve, and thus we assume Charlie is just Eve.

3. Local subsystems measurement. For each window, Alice and Bob first measure Charlie's instrument $\mathcal{L}$ to see whether there is a click in this time window, and then  measure the subsystems $J$ to learn the information that which kind of window it belongs to. After this, Alice and Bob shall know the real value of the number of effective $\zeta$ windows, $\tilde{n}_{\zeta}$, where $\zeta=\mathcal{O},\mathcal{B},\mathcal{Z}$. For the effective $\mathcal{Z}$ windows, Alice and Bob then measure the subsystems $I$ in the basis $\{\ket{01},\ket{10}\}$ to learn the bit value information of untagged bits. For the effective $\mathcal{O}$ windows, Alice and Bob assign bits $0$ respectively, and for the effective $\mathcal{B}$ windows, Alice and Bob assign bits $1$ respectively. After all this, Bob flip all his bits. Alice and Bob get raw-key strings $Z_A$ and $Z_B$ respectively. 

4. Parameter estimation. According to the values of $\tilde{n}_{\zeta}$, Alice and Bob calculate the upper bound of phase-flip error rate, and we denote this upper bound by $\bar{\tilde{e}}^{ph}$.

5. Data post processing. Alice and Bob perform the error correction and privacy amplification to raw key strings according to the key rate formula and get the final keys. 

Obviously, if Alice and Bob measure their local subsystems before they send the subsystem $S$ to Charlie, the state preparation process in the equivalent entanglement protocol is the same as that of the virtual perfect protocol. The other difference between the equivalent entanglement protocol and the virtual perfect protocol is: in the virtual perfect protocol, the parameter estimation step is after error correction, but in the equivalent entanglement protocol it is the other way around. The security can be understood in the following ways: 

In the virtual perfect protocol, suppose Alice and Bob get the real values $\tilde{n}_{\zeta}$ through a secret channel before error correction and get $\bar{\tilde{e}}^{ph}$. Then Alice and Bob perform the error correction and privacy amplification to get the final keys. In this way, the virtual perfect protocol are totally equivalent to the equivalent entanglement protocols. Alternatively, after error correction, Alice and Bob can get the estimate of $\tilde{n}_{\zeta}$, ${n}_{\zeta}$, which are used to calculate $\bar{e}^{ph}$. Since under the condition that the error correction process is perfect (or else the protocol is failed and the security coefficient of such a case has been included in $\varepsilon_{cor}$), $n_{\zeta}=\tilde{n}_{\zeta}$ and thus $\bar{e}^{ph}=\bar{\tilde{e}}^{ph}$, we can use the values of $n_{\mathcal{Z}}$ and $\bar{e}^{ph}$ to calculate the key rate.

For clarify, we shall only concern the case of the equivalent entanglement protocol and assume $n_{\zeta}=\tilde{n}_{\zeta}$ in what follows.

In the equivalent entanglement protocol, in principle, Alice and Bob can know the exactly values of the phase-flip error rate, only if they measure the subsystems $I$ in the basis $\{\ket{X_+}=\frac{1}{\sqrt{2}}(\ket{01}+\ket{10}), \ket{X_-}=\frac{1}{\sqrt{2}}(\ket{01}-\ket{10})\}$ instead of $\{\ket{01},\ket{10}\}$ for the effective time windows. A phase error occurs when Alice and Bob measure the state $\ket{X_+}$. We shall use this property to derive the formulas of calculating $\bar{{e}}^{ph}$.

In collective attack, Charlie interacts with each input state independently and identically which results that all time windows are independent and identically distributed (iid), thus we only need to consider the calculation of one window. 

We denote Charlie's ancilla state by $\ket{\kappa}$ which includes the instrument space $\mathcal{L}$. Without loss of generality, before Alice and Bob observe the measurement results and measure their local subsystems, the state shared by Alice, Bob and Charlie is
\begin{equation}
\ket{\Psi_{ABC}}=\hat{M}_C\ket{\phi}\otimes \ket{\kappa},
\end{equation}      
where $\hat{M}_C$ is a unitary operator. We denote $\xi$ as the measurement outcome of Alice and Bob. We denote $\hat{M}_D=\{\oprod{\mbox{click}}{\mbox{click}},\oprod{\mbox{no}}{\mbox{no}}\}$ as the measurement operator to the instrument space $\mathcal{L}$ where $\oprod{\mbox{click}}{\mbox{click}}$ represents the right detector clicking and $\oprod{\mbox{no}}{\mbox{no}}$ represents other measurement results. We denote $\hat{M}_L=\{\oprod{\mathcal{O}}{\mathcal{O}}\otimes \oprod{2}{2},\oprod{\mathcal{B}}{\mathcal{B}}\otimes \oprod{2}{2},\oprod{\mathcal{Z}}{\mathcal{Z}}\otimes \oprod{X_+}{X_+},\oprod{\mathcal{Z}}{\mathcal{Z}}\otimes \oprod{X_-}{X_-}\}$. 

In each time window, the probability that it causes the right detector clicking and Alice and Bob learns it is a $\mathcal{O}$ window is
\begin{equation}
\begin{split}
\Pr(\xi=\mathcal{O},\mbox{click})=&\tr(\oprod{\mathcal{O}}{\mathcal{O}}\otimes\oprod{2}{2}\otimes \oprod{\mbox{click}}{\mbox{click}} \hat{M}_C\oprod{\phi}{\phi}\otimes \oprod{\kappa}{\kappa}(\hat{M}_C)^\dagger)\\
=&\sum_{\bar{i}}\left| \bra{\bar{i}}\bra{\mathcal{O}}\bra{2}\bra{\mbox{click}}\hat{M}_C\ket{\phi}\ket{\kappa}\right| ^2\\
=& p_0^2 \sum_{\bar{i}}\left| \bra{\bar{i}} \bra{\mbox{click}}\hat{M}_C\ket{00}\ket{\kappa}\right| ^2,
\end{split}
\end{equation}
where $\{\ket{\bar{i}}\}$ is a basis for Charlie's remaining space after excluding the instrument space $\mathcal{L}$. Note that the subsystems $S$ belongs to Charlie after Alice and Bob send out them.

Similarly, the probability that each time window causes the right detector clicking and Alice and Bob learns it is a $\mathcal{B}$ window is
\begin{equation}
\begin{split}
\Pr(\xi=\mathcal{B},\mbox{click})= p_x^2 \sum_{\bar{i}}\left| \bra{\bar{i}} \bra{\mbox{click}}\hat{M}_C\ket{\mu_A\mu_B}\ket{\kappa}\right| ^2,
\end{split}
\end{equation}
and the probability that each  time window causes the right detector clicking and Alice and Bob learns it causes phase error is
\begin{equation}
\begin{split}
\Pr(\xi=X_+,\mbox{click})= \frac{p_0p_x}{2} \sum_{\bar{i}}\left| \bra{\bar{i}} \bra{\mbox{click}}\hat{M}_C(\ket{{0\mu_B}}+\ket{{\mu_A0}})\ket{\kappa}\right| ^2,
\end{split}
\end{equation} 

As shown in Ref.~\cite{wang2019practical,jiang2023side}, we can decompose $\ket{{0\mu_B}}+\ket{{\mu_A0}}$ as
\begin{equation}\label{eqalpha}
\ket{{0\mu_B}}+\ket{{\mu_A0}}=c_0\ket{00}+c_1\ket{\mu_A \mu_B}+\bar{c}_2\ket{\phi_2},
\end{equation}
where
\begin{equation}
\bar{c}_2\ket{\phi_2}=\ket{{0\mu_B}}+\ket{{\mu_A0}}-c_0\ket{00}-c_1\ket{\mu_A \mu_B}.
\end{equation}
And when $c_0c_1=1$, we have
\begin{equation}\label{c2upper}
\bar{c}_2^2=\left(c_0+c_1-2e^{-\mu_A/2}\right)\left(c_0+c_1-2e^{-\mu_B/2}\right). 
\end{equation}
The values of $c_0,c_1$ are highly arbitrary, and can be optimized to get the highest key rate. Based on experience, without losing too much performance, we can take $c_0=e^{-(\mu_A+\mu_B)/4},c_1=e^{(\mu_A+\mu_B)/4}$

Note that
\begin{equation}
\sum_{\bar{i}}\left| \bra{\bar{i}} \bra{\mbox{click}}\hat{M}_C\ket{\phi_2} \ket{\kappa}\right| ^2\le 1,
\end{equation}
we have~\cite{wang2019practical}
\begin{equation}
\begin{split}
&\Pr(\xi=X_+,\mbox{click})\le\frac{p_0p_x}{2}\left[\frac{c_0^2}{p_0^2}\Pr(\xi=\mathcal{O},\mbox{click})+\frac{c_1^2}{p_x^2}\Pr(\xi=\mathcal{B},\mbox{click})+(\bar{c}_2)^2 \right. \\
&+ \frac{2c_0c_1}{p_0p_x}\sqrt{\Pr(\xi=\mathcal{O},\mbox{click})\Pr(\xi=\mathcal{B},\mbox{click})}+\left. \frac{2c_0{\bar{c}_2}}{p_0}\sqrt{\Pr(\xi=\mathcal{O},\mbox{click})}+\frac{2c_1{\bar{c}_2}}{p_x}\sqrt{\Pr(\xi=\mathcal{B},\mbox{click})} \right],
\end{split}
\end{equation}

Denote $\mean{n_{\zeta}}=N\Pr(\xi=\zeta,\mbox{click})$ for $\zeta=\mathcal{O},\mathcal{B}$ as the expected value of $n_{\zeta}$. Since the collective attack is considered here, we can use the Chernoff bound to estimate the upper bound of $\mean{n_{\zeta}}$ according to $n_{\zeta}$. Then we have
\begin{equation}
\mean{N^{ph}}\le \mean{\bar{N}^{ph}}=\frac{p_0p_x}{2}\left[\frac{c_0^2}{p_0^2}\mean{n_{\mathcal{O}}}^U+\frac{c_1^2}{p_x^2}\mean{n_{\mathcal{B}}}^U+\bar{c}_2^2N+ \frac{2c_0c_1}{p_0p_x}\sqrt{\mean{n_{\mathcal{O}}}^U\mean{n_{\mathcal{B}}}^U}+ \frac{2c_0\bar{c}_2}{p_0}\sqrt{N\mean{n_{\mathcal{O}}}^U}+\frac{2c_1\bar{c}_2}{p_x}\sqrt{N\mean{n_{\mathcal{B}}}^U} \right],
\end{equation}
where $\mean{N^{ph}}=N\Pr(\xi=X_+,\mbox{click})$ is the expected value of the number of phase errors in the effective $\mathcal{Z}$ windows and $\mean{\bar{N}^{ph}}$ is its upper bound. Again, since the collective attack is considered here, we can use the Chernoff bound to estimate the upper bound of the real value of phase errors according to $\mean{\bar{N}^{ph}}$, and we have
\begin{equation}
\bar{N}^{ph}=O^U(\mean{\bar{N}^{ph}}),
\end{equation}  
where $O^U(x)$ is defied in Eq.~\eqref{OU}. The upper bound of the real value of the phase-flip error rate is
\begin{equation}
\bar{e}^{ph}=\bar{N}^{ph}/n_{\mathcal{Z}}.
\end{equation}


With all those, we can calculate $\bar{e}^{ph}$ according the values of $n_{\zeta}$.

\subsection{Chernoff bound}\label{chernoff}
The Chernoff bound can help us estimate the expected value from their observed values~\cite{chernoff1952measure}. Let $X_1,X_2,\dots,X_n$ be $n$ independent random samples, detected with the value 1 or 0, and let $X$ denote their sum satisfying $X=\sum_{i=1}^nX_i$. $E$ is the expected value of $X$. We have
\begin{align}
\label{EL}E^L(X)=&\frac{X}{1+\delta_1(X)},\\
\label{EU}E^U(X)=&\frac{X}{1-\delta_2(X)},
\end{align}
where we can obtain the values of $\delta_1(X)$ and $\delta_2(X)$ by solving the following equations
\begin{align}
\label{delta1}\left(\frac{e^{\delta_1}}{(1+\delta_1)^{1+\delta_1}}\right)^{\frac{X}{1+\delta_1}}&=\xi,\\
\label{delta2}\left(\frac{e^{-\delta_2}}{(1-\delta_2)^{1-\delta_2}}\right)^{\frac{X}{1-\delta_2}}&=\xi,
\end{align}
where $\xi$ is the failure probability.

Besides, we can use the Chernoff bound to help us estimate their real values from their expected values. Similar to Eqs.~\eqref{EL}- \eqref{delta2}, the observed value, $O$, and its expected value, $Y$, satisfy 
\begin{align}
\label{OU}&O^U(Y)=[1+\delta_1^\prime(Y)]Y,\\
\label{OL}&O^L(Y)=[1-\delta_2^\prime(Y)]Y,
\end{align}   
where we can obtain the values of $\delta_1^\prime(Y,\xi)$ and $\delta_2^\prime(Y,\xi)$ by solving the following equations
\begin{align}
\left(\frac{e^{\delta_1^\prime}}{(1+\delta_1^\prime)^{1+\delta_1^\prime}}\right)^{Y}&=\xi,\\
\label{endd}\left(\frac{e^{-\delta_2^\prime}}{(1-\delta_2^\prime)^{1-\delta_2^\prime}}\right)^{Y}&=\xi.
\end{align}

\subsection{Some possible variants of improved method for SCS protocols}
1. In our improved method in the main text, we have ignored the data of time windows heralded by left detector, which have large bit-flip error. We can choose to add this part with AOPP method~\cite{xu2020sending} to reduce its bit-flip rate, if it contributes positively. Here Charlie shall announce the specific detector that clicks at every time window.

2. Post selection without active phase compensation. Here, Charlie is supposed to announce an effective time windows if it has the correct phase difference of two sides and the heralded detector is not at the preferred side.

Remark: Charlie measures phase difference of two sides for all time windows. If Charlie finds the value satisfying certain conditions, e.g. around $0$ or $\pi$, for a $\mathcal B$ window, the major light beam is supposed to prefer a specific side after passing through the BS. 

This method simplifies Charlie’s action in saving active phase compensation, in the price of decreasing the key rate significantly.

\end{document}